\documentclass[aps,pra,showpacs,groupedaddress,superscriptaddress,a4paper,10pt,twocolumn]{revtex4}
\usepackage{graphicx}
\usepackage{latexsym}
\usepackage{makeidx}
\usepackage{amsmath}
\usepackage{amssymb}
\usepackage{graphicx}

\input{tcilatex}
\begin{document}

\title{Photon generation through decoherence in cavity QED: analytical
analysis}
\author{T. Werlang}
\affiliation{Departamento de F\'{\i}sica, Universidade Federal de S\~{a}o Carlos, P.O.
Box 676, S\~{a}o Carlos\textit{, }13565-905, S\~{a}o Paulo\textit{, }Brazil}
\author{A. V. Dodonov}
\affiliation{Departamento de F\'{\i}sica, Universidade Federal de S\~{a}o Carlos, P.O.
Box 676, S\~{a}o Carlos\textit{, }13565-905, S\~{a}o Paulo\textit{, }Brazil}
\author{E. I. Duzzioni}
\affiliation{Instituto de F\'{\i}sica, Universidade Federal de Uberl\^{a}ndia, Av. Jo\~{a}%
o Naves de \'{A}vila, 2121, Santa M\^{o}nica, 38400-902, Uberl\^{a}ndia, MG,
Brazil}
\author{C. J. Villas-B\^{o}as}
\affiliation{Departamento de F\'{\i}sica, Universidade Federal de S\~{a}o Carlos, P.O.
Box 676, S\~{a}o Carlos\textit{, }13565-905, S\~{a}o Paulo\textit{, }Brazil}

\begin{abstract}
The Rabi Hamiltonian describes the interaction between a two-level atom and
a single mode of the quantized electromagnetic field. When the system is
subject to the Markovian atomic dephasing reservoir, the anti-rotating term
leads to the photon generation from vacuum. In the presence of Markovian
damping reservoirs, the asymptotic mean photon number is higher than the
thermal photon number expected in the absence of the anti-rotating term. We
obtain approximate analytical expressions in the asymptotic regime for the
photon creation rate in the pure dephasing case and the mean photon number
and the atomic population inversion in the general case. Our analytical
results are valid in the small mean photon number limit and they were tested
by numerical analyses.
\end{abstract}

\maketitle

The Rabi Hamiltonian (RH) \cite{Rabi} is the simplest and the most used
Hamiltonian deduced from first principles to describe the interaction
between a two-level atom and a single mode of the quantized electromagnetic
(EM) field \cite{Scully}. It reads $(\hbar =1)$%
\begin{equation}
H=\omega n+\frac{\omega _{0}}{2}\sigma _{z}+\sqrt{2}gx\sigma _{x},
\label{RH}
\end{equation}%
where $\omega $ and $\omega _{0}$ are the cavity and atomic transition
frequencies, respectively, and $g$ is the coupling constant. The atomic
operators are%
\begin{equation*}
\sigma _{x}=\sigma _{+}+\sigma _{-},\quad \sigma _{y}=\frac{\sigma
_{+}-\sigma _{-}}{i},\quad \sigma _{z}=\left\vert e\right\rangle
\left\langle e\right\vert -\left\vert g\right\rangle \left\langle
g\right\vert ,
\end{equation*}%
where $\sigma _{+}=\left\vert e\right\rangle \left\langle g\right\vert $ and 
$\sigma _{-}=\sigma _{+}^{\dagger },$ with $\left\vert g\right\rangle $ and $%
\left\vert e\right\rangle $ denoting the ground and excited states,
respectively. The cavity field quadrature operators are%
\begin{equation*}
x=\frac{a+a^{\dagger }}{\sqrt{2}},\quad p=\frac{a-a^{\dagger }}{\sqrt{2}i},
\end{equation*}%
where $a$ ($a^{\dagger }$) is the annihilation (creation) operator of the EM
field and $n=a^{\dagger }a$ is the photon number operator.

The exact solution of the RH is lacking to the present day, so usually one
performs the RWA \cite{Scully,Schleich}, by which the so called
anti-rotating term $g\left( a^{\dagger }\sigma _{+}+a\sigma _{-}\right) $ is
neglected. The resulting approximate Hamiltonian is known as the
Jaynes-Cummings Hamiltonian (JCH) \cite{JC} and has an elegant exact
solution \cite{Scully,Schleich,Shore,klimov}, which led to a prediction of a
variety of purely quantum phenomena, such as collapse and revival of the
atomic inversion \cite{granular}, Rabi oscillations \cite{rabio}, squeezing 
\cite{8}, non-classical states, such as the Schr\"{o}dinger cat-like state 
\cite{sc} and Fock states \cite{fs}, and the atom-atom or atom-field
entanglement \cite{entanglement}. Many of these phenomena were observed
experimentally in the last decades, thereby giving an experimental
validation test of the RWA. Nowadays, JCH is the main analytic tool to
analyse cavity quantum electrodynamics (QED) \cite{Shore,haroxo}, including
protocols for quantum information processing in systems composed by an
effective two-level atom (qubit) and photons or phonons. As examples, one
can cite the already implemented architectures, such as circuit QED \cite%
{s3,PRA75-032329,Nori,revvv}, cavity QED \cite{haroxo,S298-1372} and trapped
ions \cite{Wineland}, or novel proposals, such as polar molecules coupled to
stripline resonators \cite{NP2-636}, mechanical resonators coupled to an
electronic spin qubit \cite{lukin}, etc.

The range of validity of RWA has been studied for a long time and the common
sense \cite{Scully,Schleich,klimov} is that it is valid for a weak field
amplitude, small $g$ and small detuning $\left\vert \Delta \right\vert \ll 1$%
, where $\Delta \equiv \omega _{0}-\omega $. Several numerical studies
exemplified the deviation of the dynamics of the RH compared to the one
expected from the JCH \cite{klimov,v1,v2,v3,v4}, thereby demonstrating the
breakdown of RWA in specific regimes. Alternative approximations have been
also suggested in order to increase the validity of the RWA \cite%
{Pereverzev,Irish}. Moreover, it was shown recently that the antirotating
term is responsible for the photon generation from vacuum due to an analog
of the Dynamical Casimir Effect \cite{book} in non-stationary cavity QED
systems \cite{s6,d2,mme}.

The majority of the previous studies on the role of the anti-rotating term
was performed in the idealized closed system approach, when the atom-cavity
system is isolated from the environment. Although the dynamics of the JCH
has been also extensively studied in the presence of dissipative
environments \cite{PRA75-032329,Briegel,b1,b2}, the RH has not received much
attention in open system dynamics until recently \cite{werlang}. However, it
seems that novel and unexpected phenomena appear when one combines the
anti-rotating term with the dissipation induced by several kinds of
environments \cite{werlang}.

Recently, we studied numerically the dynamics of the cavity QED system \cite%
{werlang}, described by the RH, subjected to the action of damping and
dephasing reservoirs. Under the standard Born-Markovian and the weak
system-reservoir coupling approximations \cite{s1,vvv,Briegel}, the system
dynamics is governed by the master equation%
\begin{equation}
\frac{\partial \rho }{\partial t}=-i\left[ H,\rho \right] +\mathcal{L}\left(
\rho \right) ,  \label{mastereq}
\end{equation}%
where $H$ is the RH (\ref{RH}) and the dissipation superoperator $\mathcal{L}%
\left( \rho \right) $ is given by 
\begin{equation}
\mathcal{L}\left( \rho \right) =\mathcal{L}_{a}\left( \rho \right) +\mathcal{%
L}_{f}\left( \rho \right) +\mathcal{L}_{da}\left( \rho \right) +\mathcal{L}%
_{df}\left( \rho \right) ,  \label{Lindblad}
\end{equation}%
with the standard definitions \cite{s1} 
\begin{gather*}
\mathcal{L}_{a}\left( \rho \right) =\gamma (n_{t}+1)\mathcal{D}[\sigma
_{-}]+\gamma n_{t}\mathcal{D}[\sigma _{+}] \\
\mathcal{L}_{f}\left( \rho \right) =\kappa (n_{t}+1)\mathcal{D}[a]+\kappa
n_{t}\mathcal{D}[a^{\dagger }] \\
\mathcal{L}_{da}\left( \rho \right) =\frac{\gamma _{ph}}{2}\mathcal{D}%
[\sigma _{z}],\quad \mathcal{L}_{df}\left( \rho \right) =\frac{\Gamma _{ph}}{%
2}\mathcal{D}[n].
\end{gather*}%
Above we used the short notation \cite{PRA75-032329} for the Lindblad
superoperator $\mathcal{D}[\hat{L}]\rho \equiv \left( 2L\rho L^{\dagger
}-L^{\dagger }L\rho -\rho L^{\dagger }L\right) /2$. The superoperators $%
\mathcal{L}_{a}\left( \rho \right) $ and $\mathcal{L}_{f}\left( \rho \right) 
$ describe the effects of the thermal reservoirs (with mean photon number $%
n_{t}$) on the atom and the field, respectively, where $\gamma $ ($\kappa $)
is the atom (cavity) relaxation rate. Another source of decoherence are the
phase damping reservoirs acting on the atom (field), represented by $%
\mathcal{L}_{da}\left( \rho \right) $ ($\mathcal{L}_{df}\left( \rho \right) $%
), where $\gamma _{ph}$ ($\Gamma _{ph}$) is the atomic (cavity) pure
dephasing rate. Usually, the cavity dephasing is small compared to other
dissipative channels in circuit QED, so it is neglected \cite{N445-515}.
However, due to the measurement back-action \cite{davido}, the effective
cavity dephasing rate can become large in microwave cavity QED, in which the
field is continuously measured via quantum non-demolition photon counting
using non-resonant Rydberg atoms \cite{har08}.

In \cite{werlang} we showed numerically that in the presence of only atomic
dephasing reservoir ($\Gamma _{ph}=\kappa =\gamma =0$), there is
asymptotically a linear photon growth as function of time for any initial
state $|\psi _{0}\rangle $, even for atom and field being initially in their
respective ground states, $|\psi _{0}\rangle =|g,0\rangle $. This occurs due
to the combination of the atomic dephasing and the anti-rotating term.
Moreover, for non-zero $\kappa $ and $\gamma $ the mean photon number
attains a stationary value greater than the thermal photon number $n_{t}$ in
the reservoir. The atomic population inversion also achieves a stationary
value above the one expected from the JCH. Here we evaluate analytically the
photon generation rate for the pure dephasing case and estimate the
asymptotic mean photon number and the atomic population inversion when all
the sources of loss are present in the weak coupling regime and small photon
number limit.

From (\ref{mastereq}) the Heisenberg equations of motion for the mean photon
number $\left\langle n\right\rangle $ and the atomic population inversion $%
\left\langle \sigma _{z}\right\rangle $ read%
\begin{gather}
\left\langle \dot{n}\right\rangle =-\sqrt{2}g\left\langle p\sigma
_{x}\right\rangle -\kappa \left\langle n\right\rangle +\kappa n_{t}
\label{nd} \\
\left\langle \dot{\sigma _{z}}\right\rangle =2\sqrt{2}g\left\langle x\sigma
_{y}\right\rangle -\frac{\gamma }{2}\left[ 1+\left( 2n_{t}+1\right)
\left\langle \sigma _{z}\right\rangle \right] .  \label{sd}
\end{gather}%
The equations for the higher order dynamic variables are%
\begin{eqnarray}
\left\langle \overset{\cdot }{p\sigma _{x}}\right\rangle &=&-\left\langle
x\sigma _{x}\right\rangle -\omega _{0}\left\langle p\sigma _{y}\right\rangle
-\chi \left\langle p\sigma _{x}\right\rangle -\sqrt{2}g  \notag \\
\left\langle \overset{\cdot }{x\sigma _{x}}\right\rangle &=&\left\langle
p\sigma _{x}\right\rangle -\omega _{0}\left\langle x\sigma _{y}\right\rangle
-\chi \left\langle x\sigma _{x}\right\rangle  \notag \\
\left\langle \overset{\cdot }{p\sigma _{y}}\right\rangle &=&\omega
_{0}\left\langle p\sigma _{x}\right\rangle -\left\langle x\sigma
_{y}\right\rangle -\chi \left\langle p\sigma _{y}\right\rangle -\sqrt{2}%
g\alpha  \label{2} \\
\left\langle \overset{\cdot }{x\sigma _{y}}\right\rangle &=&\left\langle
p\sigma _{y}\right\rangle +\omega _{0}\left\langle x\sigma _{x}\right\rangle
-\chi \left\langle x\sigma _{y}\right\rangle -\sqrt{2}g\zeta ,  \notag
\end{eqnarray}%
where%
\begin{equation}
\chi \equiv \gamma _{ph}+\Gamma _{ph}+\frac{\kappa }{2}+\gamma \left( n_{t}+%
\frac{1}{2}\right) .  \label{chi}
\end{equation}%
The system of equations (\ref{2}) is not closed because of the dynamic
variables $\zeta \equiv \left\langle 2x^{2}\sigma _{z}\right\rangle $ and $%
\alpha \equiv \left\langle \left( xp+px\right) \sigma _{z}\right\rangle $,
which obey the corresponding differential equations. Therefore, this system
of equations cannot be integrated exactly, although some numerical methods
for its solution based on semi-Lie algebra have been proposed \cite{v3}. For
simplicity, from now on we shall neglect the pure cavity dephasing, setting $%
\Gamma _{ph}=0$.

First, we estimate the photon generation rate for the initial state $%
|g,0\rangle $ in the absence of damping ($\kappa =\gamma =0$), under the
experimentally realistic weak coupling ($g\ll 1$) and low temperature ($%
n_{t}\ll 1$) regimes. While the mean number of photons generated through
decoherence is small, $\left\langle n\right\rangle \ll 1$, the probability
of the state $|g,0\rangle $ is high, so we assume $\zeta \simeq \left\langle
g,0|2x^{2}\sigma _{z}|g,0\right\rangle =-1$ and $\alpha \simeq \left\langle
g,0|\left( xp+px\right) \sigma _{z}|g,0\right\rangle =0$ in order to make
the system of equations (\ref{2}) solvable. To test the validity of this
assumption, we studied numerically the asymptotic values $\zeta _{a}$ and $%
\alpha _{a}$ as function of $\Delta $, $g$ and $\gamma _{ph}$ in the range
of parameters we are interested in. From now on we set $\omega =1$. As shown
in Fig. 1, $\zeta _{a}$ and $\alpha _{a}$ are always close to $-1$ and $0$,
respectively, as expected.

Thus, for $\left\langle n\right\rangle \ll 1$ the asymptotic values of the
quantities appearing on the RHS of Eqs. (\ref{nd}) and (\ref{sd}) are%
\begin{equation*}
\left\langle p\sigma _{x}\right\rangle _{a}=-\left\langle x\sigma
_{y}\right\rangle _{a}\simeq -\sqrt{2}\frac{g\gamma _{ph}}{\Delta
_{+}^{2}+\gamma _{ph}^{2}},
\end{equation*}%
where we defined $\Delta _{+}\equiv \omega +\omega _{0}$. Substituting $%
\left\langle p\sigma _{x}\right\rangle _{a}$ into Eq. (\ref{nd}) we find
that asymptotically the photon creation rate $\left\langle \dot{n}%
\right\rangle _{a}$ attains a constant value%
\begin{equation}
\left\langle \dot{n}\right\rangle _{a}\simeq 2\gamma _{ph}\frac{g^{2}}{%
\Delta _{+}^{2}+\gamma _{ph}^{2}}.  \label{ndot}
\end{equation}%
For $\gamma _{ph}\ll \Delta _{+}$, $\left\langle \dot{n}\right\rangle _{a}$
is proportional to $\gamma _{ph}$ and $g^{2}$, and inversely proportional to 
$\Delta _{+}^{2}$, as observed numerically in \cite{werlang}. In Fig. 2 we
compare the numerical values of $\left\langle \dot{n}\right\rangle _{a}$ to
the formula (\ref{ndot}) as function of $\gamma _{ph}$, $g$ and $\Delta _{+}$%
. We see that the approximate formula scales correctly with the system
parameters, but differs slightly from the numerical values due to the
deviations of $\zeta $ and $\alpha $ from $-1$ and $0$, respectively, which
were neglected in our approximation. Nevertheless, the simple formula (\ref%
{ndot}) gives the correct order of magnitude of the photon generation rate
through decoherence.

Applying the same procedure to the general case with damping, in the limit $%
\left\langle n\right\rangle \ll 1$ we obtain the following stationary values
for the mean photon number $\left\langle n\right\rangle _{\infty }$ and the
atomic population inversion $\left\langle \sigma _{z}\right\rangle _{\infty
} $%
\begin{gather}
\left\langle n\right\rangle _{\infty }\simeq n_{t}+2\Theta \frac{\chi }{%
\kappa }  \label{x} \\
\left\langle \sigma _{z}\right\rangle _{\infty }\simeq -\frac{1}{2n_{t}+1}%
+4\Theta \frac{\chi }{\gamma \left( n_{t}+1/2\right) },  \label{y}
\end{gather}%
where%
\begin{equation*}
\Theta \equiv \frac{g^{2}}{\Delta _{+}^{2}+\chi ^{2}}.
\end{equation*}%
We point out that in the special case $\omega _{0}=0$ one can easily get the 
\emph{exact} expressions for $\left\langle \dot{n}\right\rangle _{a}$ and $%
\left\langle n\right\rangle _{\infty }$, which turn out to be precisely Eqs.
(\ref{ndot}) and (\ref{x}) with the equality sign, independently of the mean
photon number. This is a consistency check of our treatment.

We can also obtain the lower bounds for $\left\langle n\right\rangle
_{\infty }$ and $\left\langle \sigma _{z}\right\rangle _{\infty }$ as
follows. The minimum values of $\left\langle n\right\rangle _{\infty }$ and $%
\left\langle \sigma _{z}\right\rangle _{\infty }$ occur in the `worst'
scenario, when the density matrix is as close to $|g,0\rangle \langle g,0|$
as possible, so $\zeta =-1$ and $\alpha =0$ hold almost exactly. Therefore,
from (\ref{x}) and (\ref{y}) we obtain $\left\langle n\right\rangle _{\infty
}\geq n_{t}+\Theta $ and $\left\langle \sigma _{z}\right\rangle _{\infty
}\geq -\left( 2n_{t}+1\right) ^{-1}+4\Theta $ and we get a simple inequality
for the stationary values%
\begin{gather}
\Theta \leq \left\langle n\right\rangle _{\infty }-n_{t}\lesssim 2\Theta 
\frac{\chi }{\kappa }  \label{q} \\
4\Theta \leq \left\langle \sigma _{z}\right\rangle _{a}+\frac{1}{2n_{t}+1}%
\lesssim 4\Theta \frac{\chi }{\gamma \left( n_{t}+1/2\right) }.  \label{w}
\end{gather}

From Eqs. (\ref{q}) and (\ref{w}) we see that asymptotically the
anti-rotating term combined with dissipative losses creates at least $\Theta
\simeq \left( g/\Delta _{+}\right) ^{2}$ photons above the thermal photon
number $n_{t}$. In the tables I and II we compare the numerical values of $%
N\equiv \left\langle n\right\rangle _{\infty }-n_{t}$ and $S\equiv
\left\langle \sigma _{z}\right\rangle _{\infty }+\left( 2n_{t}+1\right)
^{-1} $ to the lower ($N_{<},S_{<}$) and upper ($N_{>},S_{>}$) bounds given
by Eqs. (\ref{q}) and (\ref{w}), showing that the lower bound is always
satisfied. The upper bound holds in the majority of cases and is satisfied
in the order of magnitude in all the simulations we performed. The
expression (\ref{q}) could partially explain why the mean photon number
observed in \cite{har08} is slightly higher than the expected thermal photon
number -- in that case, the cavity dephasing rate $\Gamma _{ph}$ induced by
the measurement back-action can be significant, so one would expect the
generation of photons from vacuum through decoherence.

\begin{table}[t]
\begin{ruledtabular}
\begin{tabular}{ccccccc}
 $10^{3}\gamma_{ph}$  &  $10^{3}\gamma$ & $10^{3}\kappa$ & $10^{4}N$ & $10^{4}N_>$ & $10^{4}S$ & $10^{4}S_>$ \\ \hline
 $20$ & $10$ & $1$ &15.2 &  50.0 & 21.0 & 21.0 \\
 $20$ & $10$ & $3$ &19.7 &  18.0 & 28.7 & 30.0 \\
 $20$ & $10$ & $5$ &9.69 &  10.9  & 12.3 & 22.0 \\
 $20$ & $10$ & $10$ &6.50 &  6.00  & 40.0 & 37.0 \\
 $2$ & $10$ & $10$ & 2.51&  2.40  & 8.90 &10.0  \\
 $3$ & $10$ & $10$ &3.51 &3.20    & 6.00 & 13.0 \\
 $10$ & $10$ &$ 10$ & 4.51&  4.00  & 7.00 & 24.0 \\
 $20$ & $3$ & $10$ &5.78 &5.30    &48.0  &70.0  \\
 $20$ & $5$ & $10$ & 5.93  & 5.50 & 78.0&76.0 \\
\end{tabular}
\caption{Numerical values $N$ and $S$ and upper bounds ($N_{>},S_{>}$) given by inequalities (%
\ref{q})-(\ref{w}) for different values of $\gamma _{ph}$, $\gamma $ and $\kappa $ for $\Gamma _{ph}=\Delta =0$, $g =2\cdot10^{-2}$, $n_t=0$. The lower theoretical bounds are $N_{<}=10^{-4}$ and $S_{<}=4\cdot10^{-4}$.}
\label{tab1}
\end{ruledtabular}
\end{table}
\begin{table}[tbh]
\begin{ruledtabular}
\begin{tabular}{cccccc}
 $10^3g$  &  $\Delta_+$  & $10^4N$ & $10^4N_<$, $10^4N_>$ & $10^4S$ & $10^4S_<$, $10^4S_>$  \\  \hline
 $8$ & 2.0 & 0.39 &0.16, 0.38 &  1.00 & 0.64, 1.54  \\
 $10$ & 2.0 & 0.62 &0.25, 0.60 &  1.20 & 1.00, 2.40  \\
 $50$ & 2.0 & 15.7 &6.25, 15.0 &  29.0  & 25.0, 60.0  \\
 $20$ & 1.6 & 3.75 &1.56, 3.75 &  7.50  & 6.25, 15.0 \\
 $20$ & 1.4 & 4.85 &2.04, 4.90&  9.80  & 8.16, 19.6   \\
 $20$ & 1.0 & 9.58 &4.00, 9.60 &20.0    & 16.0, 38.4  \\
 $20$ & 0.8 & 14.7 &6.25, 15.0&  30.0  & 25.0, 60.0  \\
\end{tabular}
\caption{The numerical values $N$ and $S$, the lower ($N_{<},S_{<}$%
) and the upper ($N_{>},S_{>}$) theoretical bounds given by inequalities (%
\ref{q})-(\ref{w}) for different values of $g$ and $\Delta_+$ for $n_t=0$ and decay rates $(\Gamma_{ph}, \gamma_{ph}, \gamma,\kappa)=(0,0.1,1,1)\cdot 10^{-2}$.}
\label{tab2}
\end{ruledtabular}
\end{table}

In summary, we studied analytically the phenomenon of photon generation
through decoherence in the cavity QED system, described by the Rabi
Hamiltonian, coupled to Markovian dephasing and dissipative reservoirs. We
obtained a simple expression characterizing the asymptotic photon generation
rate in the pure dephasing case, which agrees with the numerical results
previously obtained in Ref \cite{werlang}. Moreover, we deduced approximate
inequalities giving the lower and the upper bounds for the asymptotic mean
photon number and the atomic population inversion when the damping is
present. These expressions were confirmed by numerical simulations, showing
that the lower bound always holds and the upper bound is satisfied in the
order of magnitude. Our expressions also agree with the exact formula in the
special case of null atomic frequency, providing one more consistency check
of our treatment. We emphasize that our results are valid in the limit of
small mean photon number, which is precisely the situation one expects in
realistic cavity QED systems. Therefore, our study demonstrates the
importance of the anti-rotating term in the open system dynamics and gives
an estimative of its influence on the experimentally observable quantities.

The authors would like to thank the Brazilian agencies CNPq (T.W. and
C.J.V-B) and FAPESP Grant No. 04/13705-3 (AVD). This work was supported by
Brazilian Millennium Institute for Quantum Information and FAPESP Grant No.
2005/04105-5.

\textbf{Figure Captions}

\textbf{Fig. 1}: Asymptotic values $\zeta _{a}$ and $\alpha _{a}$ for $%
\Gamma _{ph}=\gamma =\kappa =0$ as function of \textbf{a)} $g$ for $\gamma
_{ph}=0.05$ and $\Delta _{+}=2$; \textbf{b)} $\gamma _{ph}$ for $\Delta
_{+}=2$ and $g=0.02$; \textbf{c)} $\Delta _{+}$ for $g=0.02$ \ and $\gamma
_{ph}=0.05$.

\textbf{Fig. 2}: Comparison between the numerical values of $\left\langle 
\dot{n}\right\rangle _{a}$ (dots) and Eq. (\ref{ndot}) for $\Gamma
_{ph}=\gamma =\kappa =0$ as function of \textbf{a)} $g$ for $\gamma
_{ph}=0.05$ and $\Delta _{+}=2$; \textbf{b)} $\gamma _{ph}$ for $\Delta
_{+}=2$ and $g=0.02$; \textbf{c)} $\Delta _{+}$ for $g=0.02$ and $\gamma
_{ph}=0.05$.

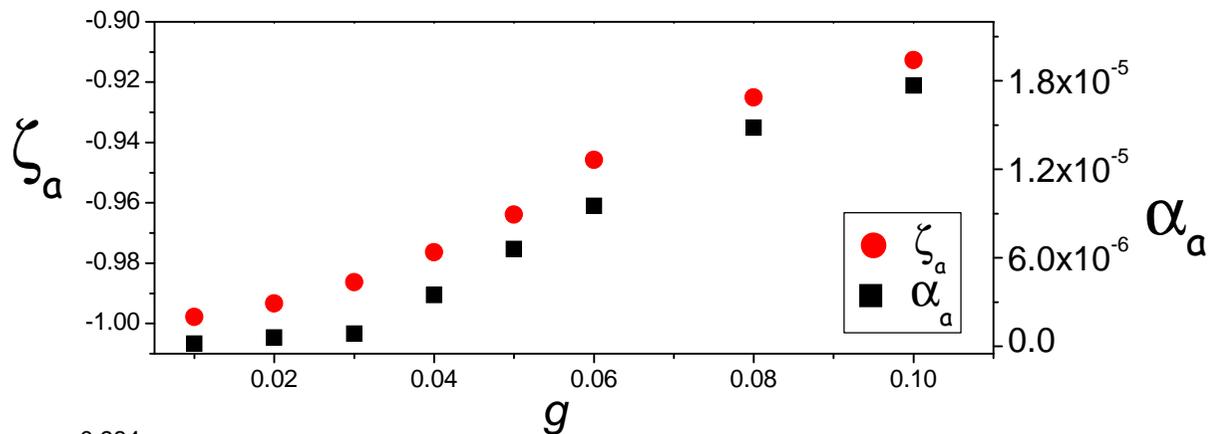
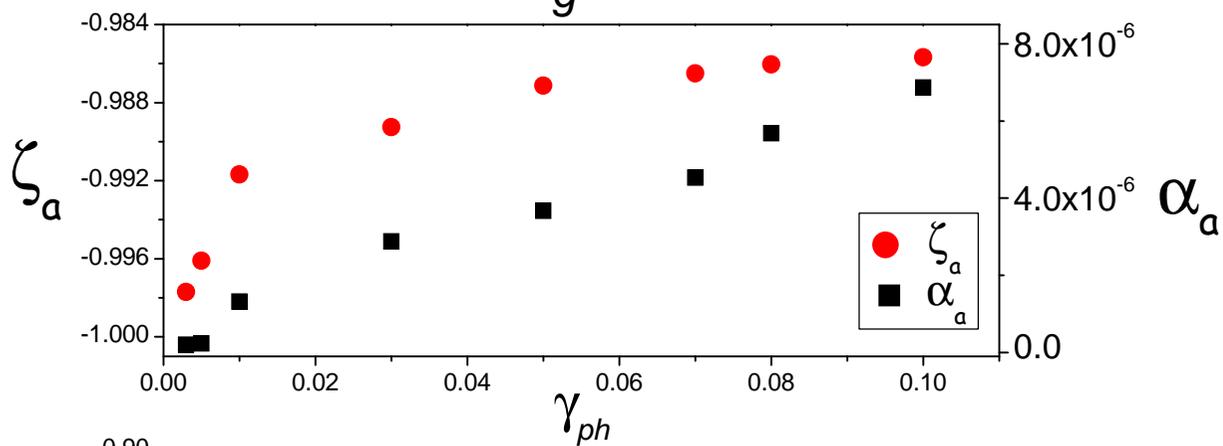
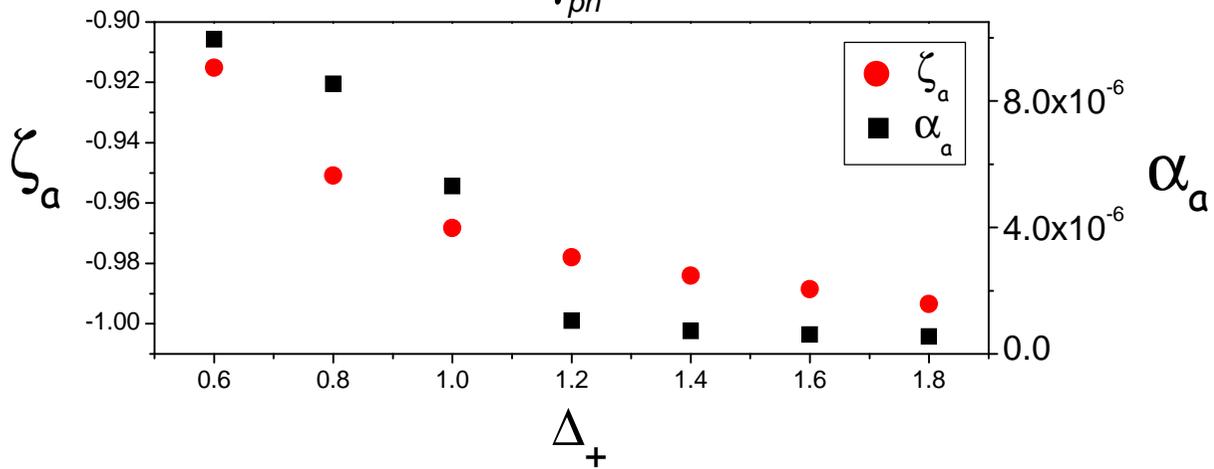
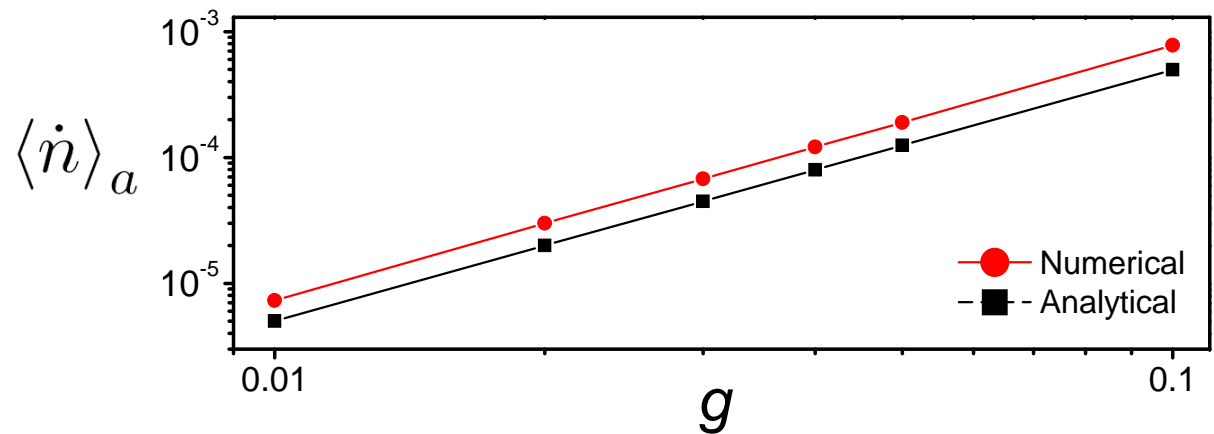
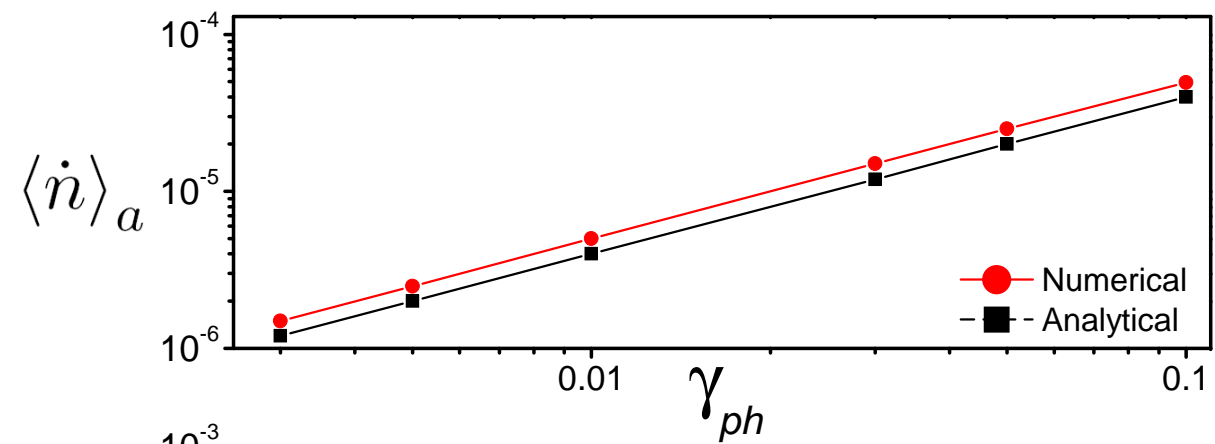
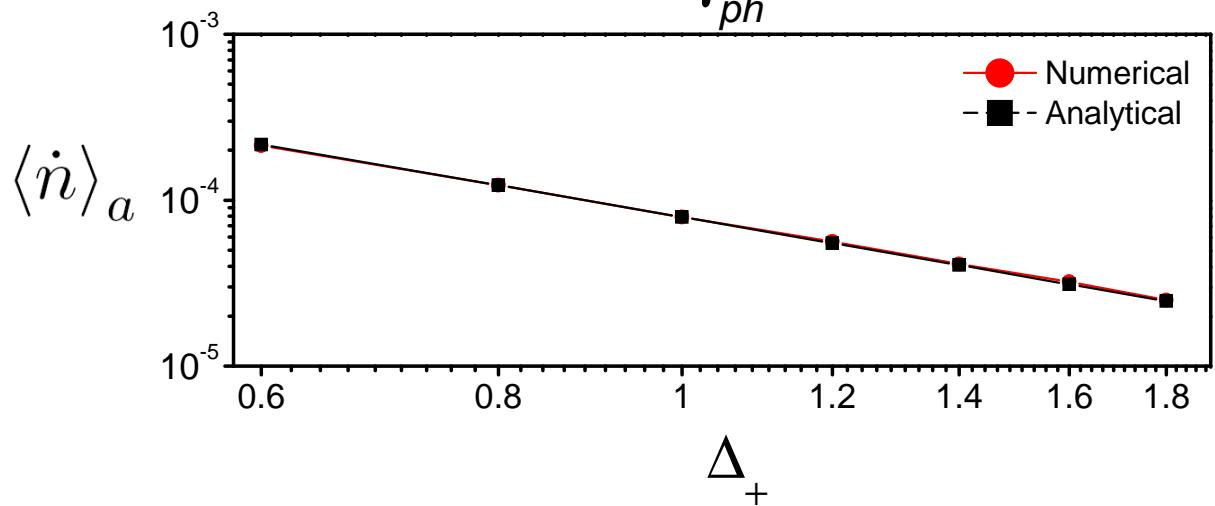
\end{document}